\documentstyle[aps,prb,multicol,floats,epsf]{revtex}
\begin{document}
\draft
\preprint{}
\wideabs{
\title{Nonlinear optical response and spin-charge separation in
one-dimensional Mott insulators}
\author{Y. Mizuno, K. Tsutsui, T. Tohyama,\cite{byline} and S. Maekawa}

\address{Institute for Materials Research, Tohoku University,
        Sendai 980-8577, Japan}
%\date{\today}
\date{Received 31 March 2000}
\maketitle

\begin{abstract}
We theoretically study the nonlinear optical response and photoexcited
states of the Mott insulators. The nonlinear optical susceptibility
$\chi^{(3)}$ is calculated by using the exact diagonalization technique
on small clusters. From the systematic study of the dependence of
$\chi^{(3)}$ on dimensionality, we find that the spin-charge separation
plays a crucial role in enhancing $\chi^{(3)}$ in the one-dimensional
(1D) Mott insulators. Based on this result, we propose a
{\it holon-doublon} model, which describes the nonlinear response in
the 1D Mott insulators. These findings show that the spin-charge
separation will become a key concept of optoelectronic devices.
\end{abstract}
}
\narrowtext

The charge gap in Mott insulators is a consequence of strong electron
correlation.  This is completely different from the band insulators,
where the charge gap is basically originated from band effects.
The nature of charge excitation across the gap is thus essentially
different between the two types of insulators. The optical response
is used to investigate the charge excitation across the gap.
In the response, the linear susceptibility with respect to the applied
electric filed, $\chi^{(1)}$, provides information on the dipole-allowed
states with odd parity.  In addition to $\chi^{(1)}$, the nonlinear
susceptibilities detect the odd states together with the dipole-forbidden
states with even parity.\cite{Butcher}

Very recently, anomalously enhanced third-order nonlinear optical
susceptibility $\chi^{(3)}$ has been reported for one-dimensional
(1D) Mott insulators of Cu oxides and Ni halides,\cite{Kishida}
as compared with those for the band insulators.  In addition,
the 1D Cu oxide, Sr$_2$CuO$_3$, exhibits ultrafast nonlinear optical
response ($\sim$1~ps) at room temperature.\cite{Ogasawara}
These facts strongly suggest a great potential of the 1D Mott
insulators as optoelectronic materials with high
performance.\cite{Kishida,Ogasawara}  Furthermore, it is shown
that $\chi^{(3)}$ in 2D Mott insulators is
smaller than that in the 1D Mott insulators.\cite{Ashida}

Motivated by these striking experiments, we theoretically examine
photoexcited states and nonlinear optical response in the Mott
insulators, and clarify underlying physics of optical nonlinearity
of the 1D Mott insulators. We use the half-filled Hubbard model
to describe the Mott insulators. The $\chi^{(3)}$ is calculated
by using the numerically exact diagonalization method on finite-size
clusters, and compared with that for higher dimensional systems
with ladder and 2D geometry.  We find that $\chi^{(3)}$ increases
with decreasing the dimensionality.  It is shown that in the
1D system dipole-allowed (odd) and -forbidden (even) states are
almost degenerate in energy, having very large dipole coupling
between the two states, while in the 2D system the dipole coupling
is rather small in spite of the closeness of the odd and even states.
We demonstrate that this peculiar feature in the 1D system is due
to the spin-charge separation inherent in the 1D Mott insulators.
We propose an effective model that can describe the optical
nonlinearity, a {\it holon}-{\it doublon} model, where {\it holon}
and {\it doublon} represent the charge degree of freedom for
photoinduced unoccupied and doubly occupied sites, respectively.
The model reproduces very well the characteristic behaviors of
the experimental $\chi^{(3)}$.\cite{Kishida,Ogasawara}
We find that the spin-charge separation becomes a key concept of
future optoelectronic devices.

The electric field {\bf E} of the incident light induces the dielectric
polarization ${\bf P}_{\rm ind}$ in a material,  which is described by
a power series of nonlinear optical susceptibility $\chi^{(n)}$:
${\bf P}_{\rm ind}=\epsilon_0\left( \chi^{(1)}{\bf E}
+ \chi^{(2)}{\bf E}^2 + \chi^{(3)}{\bf E}^3 + \cdots \right)$.
The linear susceptibility $\chi^{(1)}$ is given by
\begin{eqnarray}
&&\chi^{(1)}_{jk} (-\omega; \omega) \nonumber \\
&&=\frac{1}{\epsilon_0 L} \frac{e^2}{\hbar}\sum_a \left ( \frac{r_{0a}^j r_{a0}^k}{\Omega_a-i\Gamma_a-\omega} + \frac{r_{0a}^k r_{a0}^j}{\Omega_a+i\Gamma_a+\omega} \right),
\label{chi1}
\end{eqnarray}
where $L$ is the number of sites, $\epsilon_0$ is the dielectric
constant, $j$ and $k$ are the polarization directions, $er_{0a}$
is the dipole moment between the ground state $|0 \rangle$ and
excited state $|a \rangle$ with odd parity, $\Omega_a$ is the energy
difference between $|0 \rangle $ and $|a \rangle$, and $\Gamma_a$
is the damping factor.  Due to symmetry restrictions, $\chi^{(2)}$
vanishes in centrosymmetric materials to which Cu oxides and
Ni halides belong.  The lowest observable nonlinearity is,
therefore, $\chi^{(3)}$, which is expressed as\cite{Butcher}
\begin{eqnarray}
&&\chi^{(3)}_{jklm} (-\omega_\sigma; \omega_1, \omega_2, \omega_3)=\frac{1}{\epsilon_0 L} \frac{e^4}{3!\hbar^3} {\bf {\cal P}} \sum_{a,b,c} \nonumber \\
&&\frac{r_{0a}^j r_{ab}^k r_{bc}^l r_{c0}^m}{(\Omega_a-i\Gamma_a-\omega_\sigma)(\Omega_b-i\Gamma_b-\omega_2-\omega_3)(\Omega_c-i\Gamma_c-\omega_3)},
\label{chi3}
\end{eqnarray}
where $\omega_\sigma=\omega_1+\omega_2+\omega_3$, $b$ and $c$ denote
even and odd states, respectively, and ${\bf {\cal P}}$ represents
the sum of permutation on ($j$, $\omega_1$), ($k$, $\omega_2$),
($l$, $\omega_3$), and ($m$, $\omega_\sigma$).  In the present study,
by setting all of the polarization directions the same along chains,
we examine $\chi^{(3)}(-\omega;-\omega,\omega,\omega)$ and
$\chi^{(3)}(-\omega;0,0,\omega)$, the imaginary parts of which give
two-photon absorption (TPA) and electroabsorption spectra, respectively.
Hereafter, we set $e$, $\hbar$, and $\epsilon_0$ to be unity.

The insulating Cu oxides and Ni halides are known to be charge-transfer
(CT)-type Mott insulators, where both 3$d$ and 2$p$ orbitals exist in the
transition metal and ligand ions, respectively, and participate in the
electronic states. The values of the gap are predominantly determined
by the energy position of the $p$ orbitals. However, it is well
established that the electronic states of the CT-type insulators
can be described by the Hubbard model with single band by mapping
a bound state called the Zhang-Rice singlet state onto the lower
Hubbard band.  Actually, the single-band Hubbard model explains
very well the spectral line shape of angle-resolved photoemission and
electron-energy loss spectroscopies in the insulating
Cu-oxides.\cite{Kim1,Kim2,Neudert}

The single-band Hubbard Hamiltonian is defined as
\begin{eqnarray}
H&=&-t\sum_{\langle i,j\rangle, \sigma} \left( c_{i,\sigma}^\dagger c_{j,\sigma}+ {\rm H.c.} \right) + U\sum_i n_{i,\uparrow}n_{i,\downarrow} \nonumber \\
&+& V \sum_{\langle i,j\rangle}n_{i}n_{j},
\end{eqnarray}
where $c_{i,\sigma}^\dagger$ is the creation operator of an electron
with spin $\sigma$ at site $i$, $n_i$=$n_{i,\uparrow}$+$n_{i,\downarrow}$,
$\langle i,j\rangle$ runs over pairs on nearest neighbor sites,
$t$ is the hopping integral, $U$ is the on-site Coulomb interaction,
and $V$ is the Coulomb interaction between nearest-neighbor sites.

The realistic values of the parameters for
cuprates \cite{Kim2,Neudert,Tsutsui} are chosen: $U/t$=10, and
$V/t$=1.5 for a 1D system and $V/t$=1 for ladder and 2D systems.
Additionally, in the 2D system, a second-nearest-neighbor hopping
$t'$ is considered to simulate a parent compound of high-$T_c$
cuprates.\cite{Kim2,Tsutsui}
The damping factors are assumed to be the same for all excited
states of the systems with $\Gamma$=$0.4t$.\cite{Gamma}

We numerically examine linear absorption spectra, that are defined
by an imaginary part of $\chi^{(1)}$, and TPA spectra for a 12-site
chain with open boundary condition. The Lanczos technique is used
for the calculation of the ground state. The spectra are calculated
by using the correction vector technique.\cite{Soos}
We also calculate them for two-leg ladder, and 2D systems with 12
sites in order to clarify the dependence of $\chi^{(3)}$ on
dimensionality.  The clusters used in the calculation are shown
in Fig.~1(c).\cite{size}

The calculated results are shown in Figs.~1(a) and 1(b). The solid,
dashed, and dotted lines denote the results for the 1D, ladder
and 2D systems, respectively. The upper panel (a) represents
the linear absorption spectra, which detect odd states.
The optical gap is estimated to $\sim$2.4 (=6$t$)~eV assuming
$t$$\sim$0.4~eV, comparable with the experimental optical
gap.\cite{optical}  The linear absorption has a large magnitude
for the 1D system with strong enhancement in the low-energy region
($\omega\sim7t$), and decreases with increasing dimensionality
from the 1D to 2D systems. These results are consistent with
the experimental data of optical conductivity for 1D, ladder,
and 2D cuprates.\cite{optical}

\begin{figure}[t]
\epsfxsize=8.0cm
\centerline{\epsffile{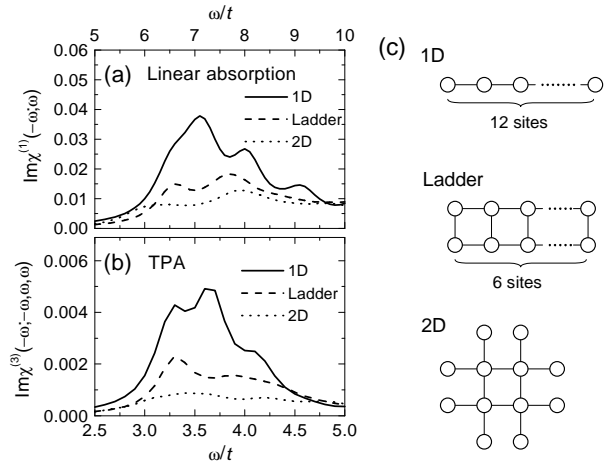}}
\vspace{2mm}
\caption{The linear absorption (a) and TPA (b) spectra for the 1D,
ladder, and 2D Hubbard models, denoted by solid, dashed, and dotted
lines, respectively. The clusters used in the calculation are
shown in (c).  $U/t$=10.  $V/t$=1.5 for the 1D cluster and $V/t$=1 for
the ladder and 2D clusters. For the 2D cluster, $t'/t$=$-$0.4.}
\label{fig:1}
\end{figure}

The lower panel (b) shows the TPA spectra, which detect even states
with the same parity as the ground state. We find that the spectral
shapes of TPA bear a close resemblance to those of $\chi^{(1)}$ when
the energy in TPA is doubled. This  means that the energy positions
of the even states lie in the same energy region as the odd states
shown in Fig.~1(a) [see the denominator of Eq.~(2)].\cite{Cu6O6}
These characteristics are independent of the dimensionality,
and agree with the results obtained by pump-probe
measurements.\cite{Ogasawara,Ashida}  We also find that the magnitude
of $\chi^{(3)}$ is the largest in the 1D system, and decreases
with increasing the dimensionality. This indicates that
one dimensionality is favorable to the enhancement of $\chi^{(3)}$,
which is consistent with the recent experimental data on Sr$_2$CuO$_3$
and Sr$_2$CuO$_2$Cl$_2$.\cite{Ashida}

In order to clarify the TPA spectra in detail, we next examine
photoexcited states and dipole moments between these states.
The dipole moments are given by $\sum_{i} \langle $odd$| x_i n_i|
$even$ \rangle $, where $|{\rm odd} \rangle$ and $|{\rm even} \rangle$
denote eigenstates with odd and even parities for the Hamiltonian
of Eq.~(3), respectively, $x_i$ is the $x$-coordinate of the position
at the $i$ site, and $n_i$ is the number operator of the electron.

Figure~2 shows the results of the energy distribution of dipole moments
for the 1D (left panel) and 2D (right panel) systems, which are
calculated exactly by using clusters shown in the insets.
The upper panel [(a) and (d)] represents the dipole moments between
the ground state and odd states. The middle panel [(b) and (e)]
shows dipole moments between the first odd state labeled as $O_1$
in (a) and (d) and even states.  In the 1D case, the $O_1$
state strongly couples to an even state with nearly the same energy
as the $O_1$ state.\cite{Cu6O6,Guo}  The magnitude of the dipole moment
is about 3.  On the other hand, in the 2D case the dipole moment
is smaller than that in the 1D system and the magnitude is less
than 0.5. In addition, other dipole moments are also very small
in the 2D system.  The lower panel [(c) and (f)] shows dipole
moments between another odd state $O_2$ and even states.
We find again that an even state with very close energy to the
$O_2$ state yields large dipole moment in the 1D case, whereas
the dipole moment in the 2D case is small although there are even
states whose energy is very close to the $O_2$ state.
These results indicate a quantitative difference of the dipole
moments between the 1D and 2D systems. The difference plays an
important role in the difference of the magnitude of TPA between
the 1D and 2D systems.

\begin{figure}[t]
\epsfxsize=8.0cm
\centerline{\epsffile{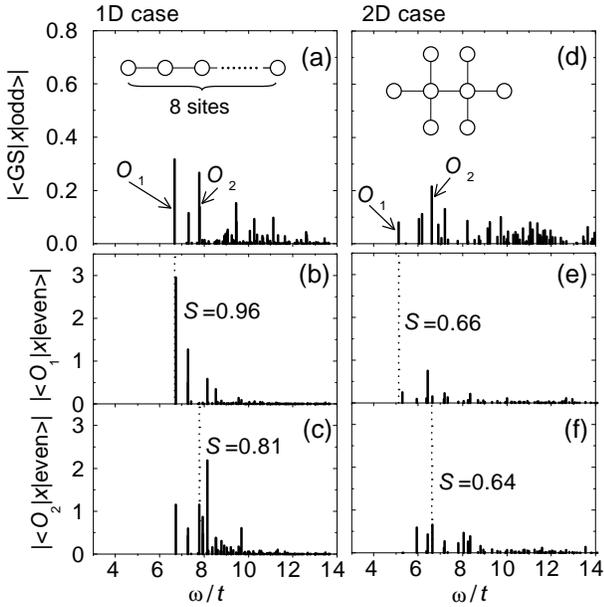}}
\vspace{2mm}
\caption{The energy distribution of the dipole moments for 8-site 1D
[(a), (b), and  (c)] and 2D clusters [(d), (e), and (f)]. The (a) and (d)
show the dipole moments between the ground state and the
dipole-allowed states. The (b) and (e) [(c) and (f)] show the dipole
moments between the dipole-forbidden states and typical two
dipole-allowed states, denoted by $O_1$ ($O_2$) shown in (a) and (d).
The ${\cal S}$ denotes a quantity to measure the similarity of the
wave functions.  $U/t$=10.  $V/t$=1.5 for the 1D cluster, and $V/t$=1
and $t'/t$=$-$0.4 for the 2D cluster.}
\label{fig:2}
\end{figure}

In general, the dipole moment becomes large when two states which
lead to the dipole moment are constructed by similar wave functions.
Therefore it is meaningful to examine the nature of the odd and
even states in the 1D and 2D systems. We define a quantity $\cal S$,
which measures the similarity of the wave functions, as follows:
${\cal S} \equiv \sum_i |\alpha_{o,i} \alpha_{e,i}|$, where
$\alpha_{o,i}$ ($\alpha_{e,i}$) is the coefficient of a basis
which constructs the odd (even) eigenstates of the Hamiltonian
of Eq.~(3), and $i$ is the index of the bases.
The quantity $\cal S$ becomes unity when two eigenstates are
the same except for the phase of each basis.

In the 1D system, $\cal S$ between the $O_1$ ($O_2$) state and
an adjacent even state is 0.96 (0.81), which is close to unity.
On the other hand, in the 2D system, $\cal S$ is 0.66 and 0.64 for
the $O_1$ and the $O_2$ states, respectively.  These results show
that in the 1D system the two wave functions with close energies
are quite similar to each other in contrast to the 2D ones.

Next, we consider the origin of the different behavior of the
photoexcited states between the 1D and 2D systems. In the 1D system,
a photoinduced carrier separates into two collective modes carrying
the spin and charge degrees of freedom (spin-charge separation).
In addition, the wave function itself is factorized as the products
of spin and charge wave functions when $U$ is very large.\cite{Ogata}
Therefore, it is reasonable in the  1D system\cite{Stephan} to
take into account only the charge degree of freedom of electrons
for the optical response.  We propose below an effective model
to describe the nonlinear optical response in the 1D system.
In contrast to the 1D system, the motion of the carriers in the
2D system is known to be strongly influenced by spin background
as long as the excitation energy is low,\cite{Tohyama} implying
that the wave function cannot be simply factorized unlike the 1D system.
It is, therefore, natural to suppose that the spin degree of freedom
in the 2D system plays a crucial role in the broad spectral weight
of $\chi^{(1)}$ and $\chi^{(3)}$ (Fig.~1) as well as the broad
distribution of the photoexcited states (Fig.~2).

Based on the spin-charge separation picture, we introduce a
{\it holon}-{\it doublon} model  for the 1D Mott insulators.
The Hamiltonian is given by
\begin{eqnarray}
H&=&t\sum_i \left( h_i^\dagger h_{i+1} - d_i^\dagger d_{i+1} +{\rm H.c.} \right) - V \sum_{\langle i,j\rangle} h_{i}^\dagger h_{i} d_{j}^\dagger d_{j} \nonumber \\
&+&\frac{U}{2}\sum_i \left(h_i^\dagger h_i+ d_i^\dagger d_i \right),
\end{eqnarray}
with the constraint of no double occupation at site $i$,
$(h_{i}^\dagger h_{i} + d_{i}^\dagger d_{i})\leq 1$.  The holon
$h$ (doublon $d$) represents the charge degree of freedom of
unoccupied (doubly occupied) sites produced by the photoexcitation.
We note that only two particles, i.e., one holon and one doublon,
are contained in the present system.  Although the model includes
the attractive Coulomb interaction, we note that the model is
different from a model for standard exciton of 1D semiconductors,
since the holon and the doublon cannot occupy the same site
due to the hard core constraint unlike the semiconductors.
As will be shown below, the constraint plays an important role in
the degeneracy of the even and odd photoexcited states.

It is easily shown in the holon-doublon model that the even and odd
states are degenerate and thus the overlap integral $\cal S$ between
them is unity: Let us introduce a set of the bases with odd and even
parities written as $|{\rm basis},o \rangle$=($|hd00\ldots\rangle
-|00\ldots dh\rangle$)$/\sqrt{2}$ and $|{\rm basis},
e\rangle$=($|hd00\ldots\rangle+|00\ldots dh\rangle$)$/\sqrt{2}$,
where $h$, $d$ and 0 denote holon, doublon, and vacant sites,
respectively.  Since there is no exchange between holon and doublon
due to the hard core constraint,
the first terms of the bases never couple to the second terms.
This means that the matrix elements of the Hamiltonian of Eq.~(4) are
independent of the sign of the second terms.  Therefore, the
Hamiltonian matrix with odd parity is equal to that with even parity.
This gives rise to the degeneracy between odd and even eigenstates
and thus ${\cal S}$=1.  This model naturally explains the
characteristics in the photoexcited states in the 1D Mott insulators
discussed above. We note that the even and odd states are always
degenerate regardless of the value of $V$ (Ref.~18) or
the range of Coulomb interactions.

\begin{figure}[t]
\epsfxsize=8.0cm
\centerline{\epsffile{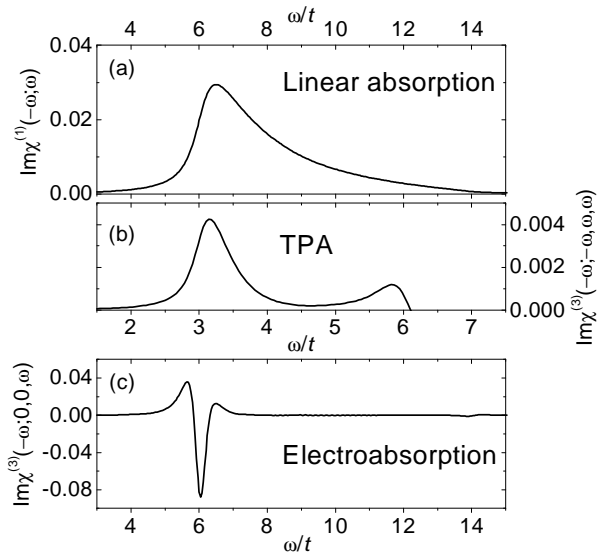}}
\vspace{2mm}
\caption{The linear absorption (a), TPA (b), and electroabsorption
(c) spectra obtained by holon-doublon model for an 80-site chain with
open boundary condition. $U$/$t$=10, $V$/$t$=1.5, and $\Gamma$/$t$=0.4.}
\label{fig:3}
\end{figure}

Figure~3 exhibits the linear and nonlinear absorption spectra in
the holon-doublon model for an 80-site chain.\cite{Model}
The linear absorption (a) and TPA (b)
spectra reproduce well those for the 1D Hubbard model shown in Fig.~1.
The obtained results are also very similar to those of the
experimental data in Sr$_2$CuO$_3$ (Refs.~2 and 3)
in the following points:  (i) The spectral weights of linear
absorption and TPA spectra are concentrated on the same energy
region when the energy in TPA is doubled. (ii) The electroabsorption
spectra (c) show an oscillating structure at the spectral edge of
the linear absorption ($\sim6t$). The agreement suggests that
this model is a proper one in describing the nonlinear optical
response in the 1D Mott insulators.

Finally we discuss the difference of nonlinear optical response
between 1D band insulators and Mott insulators. The photoexcitation
process in the band insulators creates an electron-hole pair bound
by their attractive Coulomb interaction (Mott-Wannier exciton).
The pair gives a series of bound states with odd and even parities
below the gap. In fact, in the 1D band insulators such as silicon
polymers\cite{Hasegawa} and Pt halides,\cite{Iwasa} it has been
observed that the odd and even states are concentrated on different
energy regions, and their difference is rather large ($\sim$1~eV).
However, such an energy difference between odd and even states
is generally unfavorable for obtaining the large dipole moments
because dipole moments are inversely proportional to their energy
difference.  In this respect, the 1D Mott insulators are
advantageous to large dipole moments since they have the nearly
degenerate odd and even states as shown in Figs.~1 and 3,
which leads to the large $\chi^{(3)}$.

In summary, we have clarified the nonlinear optical response and
photoexcited states of Mott insulators.  We found that $\chi^{(3)}$
is enhanced with decreasing the dimensionality, which is consistent
with the recent experiments. In the 1D system, the odd and even
photoexcited states are nearly degenerate in energy, and constructed
by similar wave functions. We found that these properties, which
become important for the large nonlinearity, are originated by
the spin-charge separation. We also proposed the effective model
in 1D Mott insulators. These findings imply that the concept
of the spin-charge separation is not of purely academic interest
but will show up as underlying physics of
optoelectronic devices in the near future.

The authors thank H. Okamoto, H. Kishida, M. Kuwata-Gonokami, and
Y. Tokura for valuable discussions. This work was supported by
a Grant-in-Aid for Scientific Research on Priority Areas from
the Ministry of Education, Science, Sports and Culture of Japan,
CREST and NEDO. The parts of the numerical calculation were
performed in the supercomputing facilities of ISSP,
University of Tokyo, and IMR, Tohoku University.


\begin{references}
\bibitem[*]{byline}Author to whom correspondence should be addressed.\\
E-mail address: tohyama@imr.tohoku.ac.jp
\bibitem{Butcher} See, for example, P. N. Butcher
and D. Cotter, {\it The Elements of Nonlinear Optics}
(Cambridge University Press, Cambridge, 1990).
\bibitem{Kishida} H. Kishida {\it et al.}, Nature {\bf 405}, 929 (2000).
\bibitem{Ogasawara} T. Ogasawara {\it et al.}, cond-mat/0002286 (unpublished).
\bibitem{Ashida} M. Ashida {\it et al.} (unpublished).
\bibitem{Kim1} C. Kim {\it et al.}, Phys. Rev. Lett. {\bf 77}, 4054 (1996).
\bibitem{Kim2} C. Kim {\it et al.}, Phys. Rev. Lett. {\bf 80}, 4245 (1998).
\bibitem{Neudert} R. Neudert {\it et al.}, Phys. Rev. Lett. {\bf 81},
657 (1998).
\bibitem{Tsutsui} K. Tsutsui {\it et al.}, Phys. Rev. Lett. {\bf 83},
3705 (1999).
\bibitem{Gamma} The value of damping factor is unknown.
However, the features of $\chi^{(1)}$ and $\chi^{(3)}$ discussed
in the present work do not depend on the value.
\bibitem{Soos} Z. G. Soos {\it et al.}, J. Chem. Phys. {\bf 90}, 1067 (1989).
\bibitem{size} No remarkable size dependence of $\chi^{(1)}$ and
$\chi^{(3)}$ was observed between 8- and 12-site clusters.
\bibitem{optical} See, for example, M. Imada {\it et al.}, Rev. Mod. Phys.
{\bf 70}, 1039 (1998) for 1D cuprates, T. Osafune {\it et al.},
Phys. Rev. Lett. {\bf 78}, 1980 (1997) for ladder cuprates, and
S. Uchida {\it et al.}, Phys. Rev. B {\bf 43}, 7942 (1991) for 2D cuprates.
\bibitem{Cu6O6} In Ref.~3, the dipole moments and susceptibilities of
the 1D two-band Hubbard model with 3$d$ and 2$p$ orbitals have
been examined numerically for a finite-size ring.
Their behaviors reported are similar to those of
the single-band Hubbard model discussed in the present work,
although the concept of the spin-charge separation is not involved in Ref.~3.
\bibitem{Guo} D. Guo {\it et al.}, Phys. Rev. B {\bf 48}, 1433 (1993).
\bibitem{Ogata} M. Ogata and H. Shiba, Phys. Rev. B {\bf 41}, 2326 (1990).
\bibitem{Stephan} W. Stephan and K. Penc, Phys. Rev. B {\bf 54}, R17269 (1996);
F. Gebhard {\it et al.}, Phil. Mag. {\bf 75}, 47 (1997).
\bibitem{Tohyama} T. Tohyama {\it et al.}, J. Phys. Soc. Jpn. {\bf 69},
9 (2000), and references therein.
\bibitem{bound} If $V>2t$, a bound state is formed for the total
momentum $K=0$ (Ref.~16).
\bibitem{Model} We note that the ground state of this model is a vacuum
without holon and doublon. The details of the calculation of $\chi^{(1)}$
and $\chi^{(3)}$ will be shown elsewhere.
\bibitem{Hasegawa} T. Hasegawa {\it et al.}, Phys. Rev. Lett. {\bf 69},
668 (1992).
\bibitem{Iwasa} Y. Iwasa {\it et al.}, Appl. Phys. Lett. {\bf 59},
2219 (1991).
\end{references}
\end{document}